# A Critical Assessment of Online Vs. Traditional Review Characteristics

*Completed Research Paper*

**Nelini Jayathilake**
University of Peradeniya
Peradeniya-Sri Lanka
nelinichamali@gmail.com

**Darshana Sedera**
Southern Cross University
Gold Coast-Australia
darshana.sedera@gmail.com

## Abstract

*With the expansion of internet-based platforms, social media and sharing economy, most individuals are tempted to review products or services that they consume. While useful in various ways, such online reviews raise questions about the reviewer's expertise, authenticity, intent, and the scales employed in such online reviews. There are emerging evidences that issues pertaining to online reviews have led to damaging consequences for businesses. With the intention of creating an awareness of reviewer characteristics and scales employed in online reviews, we present findings of 220 research papers. The review of literature identifies salient review and reviewer characteristics as well as the scale characteristics of traditional reviews and online reviews.*

## Introduction

Before the internet and social media era, consumers relied on and trusted dedicated 'reviewers' and 'critics' to assist them with purchasing decisions. As such, a few professional reviewers' bodies were formed to provide a uniform, standardized, and impartial mechanism to provide ratings. For example, the Michelin Star guide to restaurants and the HOTREC (Hotels, Restaurants, and Cafés in Europe) guide for hotels have been established and managed by domain experts from each of the respective areas (Schroeder 1985), established through well-established evaluative criteria (Titz et al. 2004). Such reviews minimize the reviewer's biasness and provide a fair assessment of the subject or service (Schroeder 1985). However, such reviews and review platforms were limited to a small number of professionals and professional bodies, creating communities of exclusivity (Verboord 2009). With the advent and massive proliferation of the internet and social media, the entire philosophy of providing reviews, critiques, and the utility of reviews have changed. Since the late 1990's internet is taking dominance over the traditional media (Nguyen 2006) and multiple critics have worked for local media in most of the cities in America (Moor 2013). Moor caught that only eleven American cities had full-time critic working for traditional media resulting a decrease in professional reviews in 2013 leaving the critic profession is in danger.

Advances in Information Technology (IT) have enabled firms to increasingly rely on open innovation (Tan et al. 2016; Cui et al. 2015). For example, any individual connected to a social media platform could provide reviews for products or services (Mayzlin et al. 2014). As such the online reviews, as compared to traditional reviews, have become (i) granular (Black and Kelley 2009), (ii) frequent (Ngo-Ye and Sinha 2014) (iii) includes a large number of reviews (Hu and Liu 2004) (iv) completed by regular consumers, without expert knowledge of the product or service. Verbrood (2009) brought into light, internet critics are non-professionals who spread evaluations through media settings low on institutionalization and high on word-of-mouth. Zinoman (2013) observed newspaper-based critics gain authority from their journalism education, the prestige of their publication, and the juxtaposition of their review next to serious news stories and internet critics are the opposite.

The popularity of online reviews increased heavily as the consumers were able to share their opinions easily through social media, shopping, and communication platforms (Hennig-Thurau et al. 2004). According to Review Monitoring, over 43% of U.S. internet shoppers end up noting their experience with products or services in such online platforms (Freddie 2019). Therefore, review platforms like Google, TripAdvisor, and OpenTable have now become the surrogates of expert reviews. Given the rising examples of how online reviews being unscrupulous, manipulated, and falsified (Luca and Zervas 2013) and given that 60-80% of consumers commence their 'shopping' process online, (Ramachandran et al. 2011), 'online reviews' can make a business blossom or wither. A positive review can improve fame, trust, and revenue of a business. Ghose and Ipeirotis (2011) discovered growth in the average subjectivity of reviews results an increase in sales for products. Moreover, Luca (2011) found out one-star increase in Yelp ratings was found to trigger 5–9% increase in revenue at restaurants. On the other hand, a negative review can damage the prestige and trustworthiness of a business leading to less sales and profitability. An interesting finding is that a single negative review could wither around 30 customers and people tend to halt businesses with no reviews or with too many negative reviews (Murphy 2020). While there are inherent advantages of having a large number of customers engaged in reviewing products or services in online platforms, some researchers argue that online critics lack the experience, impartiality, and expertise of a traditional reviewer (Burry et al., 2015). Such research highlights that most online customer reviews are simply based on a single experience that tends to be influenced by extraneous factors (Chen et al. 2020). The credibility of such online reviews is too questionable. With the evidence of a growing number of fake social media accounts, such reviews' credibility is less. For example, many popular reviewing sites such as TripAdvisor do not use a verification system like CAPTCHA or verifying email or SMS systems that exacerbate this issue (Biffaro 2015).

Similarly, though the experts' role in traditional review platforms is high, the critic approach is not always the best. A study carried out by Senecal and Nantel (2004) shows that the online reviewer's recommendations are most influential. However, they are perceived as possessing less expertise when compared to 'human experts' and less trustworthy than 'other consumers.' Besides, online reviewers can provide low-level details that otherwise would be absent in a professional review. For example, while a HOTREC provides a star rating for a hotel, online reviewers could provide a detailed view of the facilities and issues regularly (Gretzel and Yoo 2008). Furthermore, online platforms and the frequency of online reviews allow organizations to be better connected with their customers (Zhu and Zhang 2010). Finally, the freedom that online reviewers experience in writing comments would allow them to identify genuine issues in real-time, making it easier for organizations to improve their products or services.

The objective of this paper is to commence an informed discussion on the merits of online reviews of the novices versus the traditional reviews of the experts. Our research is motivated by the perceived biasness that online reviewers (and reviews) seem to have in relation to i) a large number of amateur online reviewers, ii) credibility of online review accounts, iii) the small number of traditional reviewers making it challenging to reach to customers and iv) issues pertaining to scales, measurements and emojis that may not be sophisticated to capture nuanced views of reviews.

## Methodology

This study aims to discuss the traditional vs. contemporary reviewer, review, and scale characteristics. As there is no previous study on this topic, it is difficult to develop a theory. Therefore, we have used literature search as the best approach. We have used exploratory methodology as it is flexible and helps to get a better understanding of characteristics of traditional and online reviews, reviewers, and scales (Vanhamme 2000). We examined existing multiple resources on this topic in order to gain familiarity with our research topic and to lay foundation for a more extensive study.

First step was to search for the scholarly articles. As there is no universal standard for article selection (Shamseer et al. 2015), we extended our search by referring multiple sources without limiting to academic discipline, publication status and region with the aim to obtain as many relevant articles as possible (Sigerson and Cheng 2018). As the first step, key words that used to locate the reports were "review", "reviewer", "characteristics" and "rating scale". After referring those words in a thesaurus (i.e., https://thesaurus.yourdictionary.com), we created a list of other keywords such as "critic", "critique", "criticism", "online reviewer" and "online review". Later, we added another set of keywords that are related to reviewer, review and scale and measurement characteristics. (i.e., "expertise", "trustworthiness", "credibility", "identity disclosure", "online attractiveness", "objectivity", "readability", "social media", "Facebook", "YouTube", "sharing economy", "scale", "rating").

We reviewed 45 peer-reviewed papers for the traditional reviewers that identified 14 characteristics. Further 14 online reviewer characteristics were identified by reviewing 90 papers after the year 2000. A limited number of research papers are available on traditional review and scale and measurements. However, we have identified 9 traditional review features reviewing 5 literature sources. We have reviewed 60 papers after the year 2000 to discover online review features. By reviewing 20 papers under scale and measurement, 12 scale and measurement features were identified. A total of 220 peer-reviewed papers were reviewed in order to set a strong foundation for our research.

## Results and Discussion

In this section, we will discuss the reviewer's characteristics that have a significant influence on the review he/she is going to make. Here, we will explore the characteristic similarities and differences between traditional and contemporary reviewers as well as review and scale characteristics.

### *Reviewer Characteristics*

According to the Cambridge Dictionary, "A reviewer is someone who writes articles expressing their opinion of a book, play, film, etc." Reviews on books, restaurants, arts, music, and movies are still prevalent in newspaper columns before the internet era. However, with the spread of the internet and technologies, online reviewing platforms began to outstretch (provenexpert.com).

Similarities and differences were identified while analyzing previous literature on traditional (offline) and an online reviewer. If a food journalist writes about restaurants and hotels, as an ethic, he/she pays several visits to the selected restaurant anonymously (Titz et al. 2004). However, he/she writes the review for the media with his/her real name (Association of Food Journalists n.d.). The real name of the critic is the most shared information. In contrast, online reviewers share their real name, nickname, location, hobbies (Forman 2007), profile picture (Forman, 2007; Karimi et al. 2017) over the online review platforms. Nevertheless, in many instances online reviewers do not disclose their identity (Kobayashi 2017) and demonstrate an abnormal behavior (Postmes and Spears 1998).

Screen names like 'missgussie' and 'dozen' withhold reviewers' identities, and even if reviewers use names that appear to be real names, readers are not sure that these names are the reviewers' actual identities (Mackiewicz 2010). It is interesting to note that in online reviewing, people keep their faith mostly on the reviews posted by the reviewers with profile images (Karimi et al. 2017). A study by Spiegel Research Center (2017) found that reviews from verified shoppers have significant implications for online sales compared to anonymous reviewers. Though there is a difference in the level of revealing reviewer's identity, it is one of the important characteristics relates to both.

Miller (1987) defines expertise as superior knowledge about an object or topic and the ability to judge critically and professional specialization. However, the definition for expertise of online reviewer varies. Some researchers define reviewers' expertise as having internet/computer skills (Yoo and Gretzel, 2009) or being experienced web users (Zhao et al. 2015). Some suggest as having a longer membership, earning a higher status badge (Gretzel et al. 2007) like the "elite" badge on Yelp and Top 10000 reviewers' badge on Amazon, and receiving more helpful votes (Ma et al., 2013; Xie and So, 2018).

Contrarily, expertise of the traditional reviewer includes years of experience (Titz et al. 2004), evaluative skills and communication skills (Weiss and Shanteau 2003), and service of analysis, illumination, and interpretation (Clarke 2017). When compared to expertise of online reviewer, traditional reviewers are equipped with good set of skills (Schroeder 1985; Barrows et al. 1989; Eliashberg and Shugan 1997) and are always socially conscious and respect ethics (Schroeder 1985). Simply, because they are professionals in the field. The Table 1. below shows the characteristics associated with traditional reviewer.

**Table 1. Traditional Reviewer Characteristics**

| Characteristic | Source |
|---|---|
| Possess a critical mind | Landry (1940); Lang (2014); Heiman (1997); Hervey (1911); Popkin (1960) |
| Pay multiple visits | Titz et al. (2004); Schroeder (1985); Goodsir et al. (2014); Lang (2014) |
| Subjectivity | Bordwell (2011); Schroeder (1985); Goodsir et al. (2014) |
| Expertise | Schroeder (1985); Barrows et al. (1989); Eliashberg and Shugan (1997) |
| Journalism Experience | Schroeder (1985); Barrows et al. (1989); Goodsir et al. (2014) |
| Credibility | Schroeder (1985); Barrows et al. (1989); Heiman (1997); Hervey (1911) |
| Identity Disclosure | Goodsir et al. (2014); Butler (2018); Maynard (2006); Burry et al (1985) |
| Possess General Standards | Theodore (1979); Bucak and Kose (2014) |
| Are socially conscious | Landry (1940); Eliot (1982); Gallagher (1985) |
| Ethical | Schroeder (1985) |
| Avoid free offers | Schroeder (1985); Goodsir et al. (2014); Hou (2012) |
| Free from bias | Schroeder (1985); Goodsir et al. (2014); Landry (1940); Hervey (1911) |
| Open to various interpretations | Schroeder (1985); Goodsir et al. (2014); Butler (2018); Searle et al. (1974) |
| Reviewers' personal factors | Schroeder (1985); Landry (1940); Searle et al. (1974); Popkin (1960) |

In the context of online reviewers, they are often capricious and circumstantial (Beaton 2018) up against professional reviewers. Professionals are known and expected to provide valuable and reliable information regarding the subject they make the review (Eliashberg and Shugan 1997). Journalists, reporters, and editors closely monitor, follow, and subscribe to tweets and Facebook comments, watch YouTube videos, and read blogs to search and source news and newsworthy information (Palekar and Sedera 2015). They can be perceived by online users as a popularity or eye-catching indicator of the corresponding product, because professional reviews, unlike abundant user-generated Word of Mouth (WOM), are normally only offered on a limited number of products by a small group of specialized experts (Zhou and Duan 2016).

It is noteworthy that some consumers rely on online recommendations rather than traditional recommendations. However, all the online review sources are not equally influential. Eliashberg and Shugan (1997) studied the movie industry. According to their study, consumer critics evaluate the movies and then write the review from a personal perspective. But, review writings by professional critics are different: instead of evaluating the movie, professional critics describe the movie. They are not going to present their perspective while writing the review. There exists a difference in review content create by the traditional and the online reviewer. Even though online reviewers are important implications for discourse on the internet (Jong and Burgers 2013).

Professional critics avoid free offers as if they accept any, the review will be a biased and unfair (Schroeder 1985; Goodsir et al. 2014; Hou 2012). Traditional critics always try to be strict to the standards (Theodore 1979; Bucak and Kose 2014) when reviewing as it affects their professional dignity. Another interesting fact to note that is traditional reviewers are open-up for more than one interpretation (Ambiguity) in opposition to most of the online reviewers (Schroeder 1985; Goodsir et al. 2014; Butler 2018; Searle et al. 1974). This can be clearly seen in book and film reviewing.

Credibility features are quite similar for both instances. Metzger et al. (2010) has introduced a scale with five dimensions: accuracy, believability, bias, completeness, and trustworthiness to measure online credibility. According to Ohanian (1990), reviewer trustworthiness incorporates the features dependable, honest, reliable, sincere, and trustworthy. Whitehead (1968) added another dimension to source credibility, which is objectivity.

Source credibility is one of the significant issues aroused with the invent of online review systems. People started to write fake reviews for themselves or to their rival firms expecting financial gains, and in order to promote or demote business organizations (Li et al. 2014). We can find plethora of researches conducted on the topic source credibility and online review platforms. Several mechanisms such as filtering algorithms to identify fake reviewers and reviews have introduced by scholars to overcome some issues associated with credibility (Luca and Zervas 2013; Metzger et al. 2010). Credibility of online reviewers are questionable as they can fake their identities as well as their opinions (Mackiewicz 2010).

However, traditional critics are considered credible, as they are skillful with a proper education journalism experience and industry experience (Brown 1978). Writing reviews for the public as part of their profession, and they have certain ethics guides on writing reviews (Association of Food Journalists n.d.). Hence, the credibility of reviews written by traditional reviewers is high compared to the contemporary reviewers.

Previous literature shows that the modern-day reviewers are mostly non-experts. They are frequently reviewing to build up their online social status and be voguish in the cyber world (Gretzel et al., 2007). Some characteristics are unique to the online reviewers such as online attractiveness (Gretzel et al. 2007; Pinch and Kesler 2011; Guo and Zhou 2016; Ma et al.2013). It is the number of friends/followers a reviewer has (Trevor and Filip 2011; Zhu et al. 2014). Some other scholars said it is the amount of peer's attention, acceptances, and emotional approvals (Zhu et al. 2014). In 1985, McGuire has introduced an attractiveness model including source's "familiarity", "likability", "similarity," and "attractiveness.". However, Zhu et al. (2014) classified the reviewer's attractiveness as one of the conceptual dimensions of source credibility. But, in this research we have excluded the online attractiveness from credibility features and categorize separately. As we have identified from the search of previous literature, characteristics of online reviewers are shown in the Table 2.

**Table 2. Online Reviewer Characteristics**

| Characteristic | Source |
|---|---|
| Identity Disclosure | Allington (2016); Liu and Park (2015); Gretzel et al. (2007) |
| Online Attractiveness | Gretzel et al. (2007); Pinch and Kesler (2011); Guo and Zhou (2016) |
| Expertise | Allington (2016); Liu and Park (2015); Gretzel et al. (2007) |
| Reputation | Liu and Park (2015); Hu et al. (2008); Pinch and Kesler (2011) |
| Credibility | Liu and Park (2015); Gretzel et al. (2007); Hu et al. (2008) |
| Reviewer Engagement | Gretzel et al. (2007); Hu et al. (2008); Chen and Huang (2013) |
| No General Standards | Kobayashi et al. (2015); Jong and Burgers (2013) |
| Biased Review Behavior | Gretzel et al. (2007); Hu et al. (2008); Pinch and Kesler (2011) |
| Volunteer | Yan et al. (2013); Berlin and Fisher (2004) |
| Mutual Indebtedness | Gretzel et al. (2007); Ren et al. (2013); Gilbert and Karahalios (2010) |
| Reviewer Exposure | Ngo-Ye and Sinha (2014) |
| Innovativeness | Goldenberg et al. (2009); Pan and Zhang (2011) |
| Personal Factors | Allington (2016); Gretzel et al. (2007); Pinch and Kesler (2011) |

Oxford English Dictionary has defined reputation as ''the relative estimation or esteem in which a person or thing is held'' (Simpson and Weiner 1989). However, it is not necessarily an accurate indication of that person's or object's qualities (Weigelt and Camerer 1988). The definition of the reputation of a website is quite different. Ghose and Ipeirotis (2011) have defined online reviewer reputation as the "identity-descriptive information displayed on review platforms for users who have contributed reviews" such as reviewer ranking, usernames, and special badges. Simply, the reviewer's ranking on the review platform is known as the reputation (Li et al. 2020). Reviewer reputation plays a vital role in online communities as the reviews from reputed consumers considered more reliable (Gretzel et al. 2007). In dianping.com, every user has a reputation level according to their contributions allowing users to easily identify other users with higher reputation levels (Jin et al. 2010).

Recency and the frequency of reviewing were categorized under overall reviewer engagement (Ngo-Ye and Sinha, 2014). Being recent and reviewing frequently helps online reviewers to win the attention among the community. Such online reviewers are provided with special promotions and badges (Gretzel et al. 2007). Therefore, continuously engaging in review activity is identified as a contemporary reviewer characteristic. Online consumers showcase a biased review behavior (Gretzel et al. 2007) and there is no proper standardization (Kobayashi et al. 2015). These reviewers voluntarily participate in online opinion platforms and they can switch to other online platforms very easily without any cost (Jin et al. 2010). However, they are innovative (Goldenberg et al. 2009) and most importantly online reviewers are mutual indebted (Ren et al. 2013).

Reviewer attribution is a common trait to both, which describes the reviewer's personality, traits, character, personal style, attitudes, and mood (Chen and Lurie 2013). The study has found that positive reviews are more attributed to the reviewer than negative reviews because they have more personal reasons such as the reviewer's motivation, traits, moods, and attributes (Gilbert and Malone, 1995) to do a positive review. For example, people sometimes post positive reviews to "look good" to themselves or others (Chen and Lurie 2013) or to feel the pleasure obtained from helping others (Jin et al. 2010).

*Review Characteristics*

Oxford dictionary defines the review as "A critical appraisal of a book, play, and film published in a newspaper or magazine". Product recommendations, reviews and lists of favorites provided by other users plays a crucial role in the selection of products (Kurnia et al. 2005). A Nielsen (2015) report finds that consumers trust recommendations or opinions from other consumers' more than traditional forms of advertising such as commercials or product placements on mass media, showing the persuasive power of online product reviews. Bright Local Survey revealed that 82% of consumers read online reviews for local businesses while 52% of consumers in the age group 18-54 'always' read online reviews (Murphy 2019). An amusing fact is that the average consumer reads ten reviews before feeling able to decide. Online shopping is a riveting experience, and digital buyers often search through various websites with heaps of different products to find what they want to buy. Here, the role of product ratings and reviews come to play, as 53% of people consider product ratings and reviews as one of the most crucial attributes of online shopping (Clement 2019). Consumers read reviews on various products and services they intend to use or buy. They look for reviews on restaurants (Yelp.com), books (Amazon.com), travel (Trip Advisor), movies (film.com), groceries (amazonFRESH.com) medicines (webmd.com), cosmetics (beautypedia.com), clothing, and accessories (ebay.com) and many more.

Online consumer reviews are one form of electronic WOM and include positive or negative statements made by the consumers about their experiences or opinions on a product or service they consumed (Jalilvand et al. 2011). Online review attributes are different in contrast to the traditional review criterion.

According to Rogers (2020), writing a quality review is an art that one should master. A good critic must learn everything that he can. Knowing the subject is the most important thing. Writing reviews by knowing a little about the subject is not a characteristic of an expert reviewer. There is a minimal number of studies that discuss the characteristics of traditional reviews. Though it is limited, previous literature under traditional criteria has identified several characteristics. They are objectivity, freedom from prejudice, relevance, identification of focus, an evaluative judgment, personal assessment, and linguistic and stylistic characteristics. Table 3 illustrates tradition review features.

**Table 3. Traditional Review Characteristics**

| Characteristic | Source |
|---|---|
| Objectivity | Drewry (1974); Kamerman (1979); Corrigan (2015) |
| Accuracy | Characteristics of a Good Book Review (1980); Drewry (1974) |
| Relevance | Characteristics of a Good Book Review (1980); Drewry (1974) |
| Critical evaluation | Characteristics of a Good Book Review (1980); Drewry (1974); (Kamerman 1979); Corrigan (2015) |
| Personal assessment | Characteristics of a Good Book Review (1980); Drewry (1974); (Kamerman 1979); Corrigan (2015) |
| Readability | Drewry (1974); Corrigan (2015) |
| Organization and Layout | Drewry (1974); Kamerman (1979) |
| Balanced | Kamerman (1979); Sorensen and Rasmussen (2004); Teitelbaum (1998); Corrigan (2015) |
| Non-Committal/ Superficial | Characteristics of a Good Book Review (1980); Drewry (1974) |

Though online review attributes provide consumers insights on online reviews, the quality or the professionalism of these online reviews is an open question. The reason is that very few professionals engage in writing online reviews, and some professionals use their professionalism to earn money by writing fake reviews.

Nevertheless, online reviews have a significant influence on consumers' purchasing decisions (Bounie et al. 2008), there are several issues associated with online reviews. Most of the online reviews include phrases like "I think" and "In my opinion." Using such phrases is always avoided by professionals. Novice critics use those words repetitively as they are afraid of writing declarative sentences (Rogers 2020). In traditional review systems, critics publish their strong opinion about the topic that they critically evaluate.

Many of the new critics on the Internet are not professional, but engage in participatory practices, or peer production, such as adding user-generated content, rating other persons' reviews, sharing content with friends, etc. (Verbrood 2009). In contemporary criteria, very few reviews written by professionals can be identified. Many reviewers are motivated to write reviews as it is paving the path for social recognition. Lacking the incentive of payment, these inexpert reviewers do not have clear motivations. Presumably, they think that their reviews, their recommendations, will receive at least a little attention from some audience (Mackiewicz 2010).

Moreover, as individuals and organizations are using online reviews in their decision-making process, opinion spamming, or writing fake reviews has become a business (Li et al. 2014). Most of the vendors write reviews for themselves or their competitors to make a bad reputation or harm the goodwill of a successful business organization, other rivalry firms will make fraud reviews over the internet (Luca and Zervas 2013). Monetary gain and incentives are the invisible hands behind the review fraud (Luca and Zervas 2013). Some businesses buy reviews from review selling sites while some pay to write fake reviews. Disgruntled customers are the ones who write fake negative reviews that may not be entirely fake (Boost Reviews Australian 2019).

These days, business organizations hire professional manipulators to write reviews while posing as consumers. There are some situations where misleading positive reviews have revealed (e.g., Todd Rutherford). Further, he stated that around 31% of consumers rely more on a product with "excellent" (5 stars) reviews. University of Illinois data mining expert Bing Liu estimates that one in three online reviews is fake and nearly impossible to discern from authentic ones (Salisbury 2018).

An online review is expected to be helpful (Ganu et al. 2013) and worked out with great care and nicety of detail (Racherla and Friske 2012). Star rating is a very popular feature included in the review (Pinch and Kesler 2011). Many scholars have identified valence (rating for a product) and volume (number of reviews available for a certain product) as some online review attributes. These attributes are listed below in Table 4. with the referred literature.

*Scale and Measurement Characteristics*

There are only a small number of studies about the scales used in reviews in the traditional setting. Many professionals have used star ratings in their reviews. Much of the traditional reviews took place using experience narratives (Anderson 1998; Blank 2006), a summary of large surveys (Blank 2006), and expert commentaries (Blank 2006; Bucak and Kose 2014). There is a plethora of studies that have discussed online rating scales and the vital role that they play in the broad context of e-commerce. For example, Netflix, Amazon, Yelp, Google, YouTube, Pandora, Trip Advisor and many other websites, apps use rating scales to personalize the individual user experience (Balboni 2020).

Researchers have discovered various rating scales. Online rating scales tend to focus on Unary options like "I like it" and Binary options like "Thumbs up/Thumbs down" which is common in YouTube (Sparling and Sen 2011). Moreover, different point rating scales are also widely used in various online platforms. A scale of 1- 5 rating is used in Amazon, TripAdvisor and Google review. 1–10 rating scale is used to rate movies by IMDb while Rotten Tomatoes website uses 1–100-point scale. However, in order to make the e-businesses more personal, business firms (e.g., intercom) drawn to use an emoji scale with a row of faces representing a five-point scale from "terrible" to "amazing". Another widely used online scale is the symbol of "Heart" which is being used to display liking to a something. Twitter,

Instagram, and e-commerce apps alike Etsy, Rent the Runway use this scale. The aim of using these shapes is to deliver more personalized experiences (Balboni 2020).

On a unary scale, there is one option. There is no way to express negative feedback. Facebook popularized it at the beginning (e.g., like) (Sparling and Sen 2011). The binary scale allows us to express likes or dislikes. It will not allow the users to express the percent or the proportion of their like/dislike. An issue with binary scale is over simplicity (Utterback 1996). The most widely used scale is the star rating and is common among e-commerce sites. It can be a three, five, seven, nine, or zero to a ten-point rating. Star rating is popular as reviewers find it easy and consumers are willing to give stars to each of the products connected and purchased (Lin and Heng 2015). They view star rating trustworthy (Aral 2014). Amazon shows that the main advantage for the customer arises from reviews and ratings of other customers as well as from recommendation systems based on collaborative filtering (Kurnia et al. 2005). A survey has found that 68% of United States internet users consider star rating when judging a brand or a retailer (Clement 2020). According to a Local Consumer Survey, star rating remains the most critical part of a review, within 54% of respondents stating this as a key consideration. Further, it has revealed that 39% of the users mention that the minimum star rating they consider is a rating of 3 out of 5 stars, and 9% of consumers will not be motivated to use a business with an average star rating of less than 5 out of 5 (Murphy 2017).

Though there are positive implications of online rating and scale, there are adverse effects as well. Research by Aral, 2014 suggests that online ratings are systematically biased and easily manipulated.

## Conclusion

In the study of traditional and contemporary reviewers and review characteristics we have identified many features, organizations need to pay more attention to. Due to the rising of online businesses, day by day the significance of online reviewer and review characteristics has increased. A study by Goh et al. (2013) shows that engagement in social media brand community leads to a significant increase in consumer purchases. Hence, business organizations should maintain a base of reviewers in order to boost their sales. On the other hand, improved WOM can instead benefit its potential competitors, which backfires to hurt its own sales (Lin and Heng 2015). Therefore, Lin and Heng suggest businesses should conscientiously monitor how the WOM evolves subsequently, paying attention to potential new reviewers. Whenever WOM threatens to decline, business owners should find marketing strategies to please these potential new reviewers. So, it is important to identify reviewer types, their attributes, review characteristics as well as scale and measurement features in order to develop the strategies.

Attributes of traditional expert reviewers identified by reviewing previous literature includes having a critical mind, paying multiple visits if you are reviewing a restaurant/hotel, subjectivity, expertise knowledge, having journalism experience, work under set of general standards, always being ethical, strictly avoid any type of freebies, free from bias, personal attribution of the reviewer and open to various interpretations-especially when reviewing books, art, drama, and movies.

Characteristics of novice online reviewers are identity disclosure, online attractiveness, expertise of the reviewer which includes mostly computer skills, online reputation, credibility, mutual indebtedness, exposure to various interpretations, reviewer engagement, not having general standard to write reviews, having a biased review behavior, doing the review activity voluntarily, showing innovativeness and the personal attribution of the reviewer.

As per our findings traditional review characteristics are objectivity, accuracy, relevance, displays a critical evaluation, based on personal assessment, readability, proper organization, and layout, balanced and is non-committal/superficial. In contrast, online review attributes include having a star rating together with the review, elaborateness, content, review valence, helpfulness, volume of the review, timeliness, review subjectivity, readability, credibility, and consistency. Using our literature research, we found ambiguity, valence, volume, variance, biasness, susceptibility, informality, simplicity, interpretation that differ from person to person, distracting, familiar symbols and enjoyability as characteristics of online scale and measurements.

However, we do not recommend viewing solely either traditional reviews, or online reviews to make purchase decisions as both criteria contain their own advantages and disadvantages. To overcome issues associated with both review types, we suggest using a mix of both traditional and contemporary review criteria, which will lead to a better purchase decision.

## Practical Implications

Online virtual community sites require a critical member mass in order to sustain in the future. For that online communities must retain helpful reviewers maintaining the critical member mass and even increasing it, which will escalate the net earnings of a business (Lee et al. 2011). Hence, online communities should encourage members to post reviews (Lee et al. 2011). Learning and investigating more about traditional and online review features lead business organizations to find new strategies to improve in areas where they require an upgrade. For instance, if they are getting fewer expert reviews, firms must find innovative ways to encourage professional reviews and helpful reviews. At the same time e-businesses and online communities must take steps to avoid negative reviews as a single negative review can drive away approximately 22% of customers, whereas around three negative reviews can drive away 59% of the customers (Murphy 2020).

Nowadays, many review websites provide social network functions that allow readers to follow reviewers or to rank reviewers according to their review quality in order to attract more users to post reviews and to make friends with each other (Cheng and Ho 2015). According to Bowman (2019) Facebook was one of the first large platforms to offer business reviews starting in 2013 and today, there are nearly 1.5 billion people who log in every day, offering a large audience for businesses. So, it is significant that business organizations must focus more on social network sites as well. Moreover, one of the emerging strategies, "technology-enabled gamification" can use to induce consumers to change their behaviors (Tan et al. 2017). This study helps the business organizations to develop marketing mechanisms according to the consumer group as this study gives an insight of characteristics of traditional and contemporary reviewers, reviews and scale and measurements separately. Bright Local Survey has found that when writing a review 20% of consumers expect to receive a response within a day. Therefore, business organizations must response to their customers within 24 hours as customers mean business (Murphy 2020).

Finally, online business firms and communities will be benefitted in the long run by investing time and resources into an online reputation management where it allows to manage negative business reviews and encourage satisfied clients to post more positive reviews (Bowman 2019).

## Limitation and Recommendation for Future Research

Generalizability is one of the main limitations of this research. Although, we have reviewed 220 reports from many sources which relates to different fields such as products, movies, books, theater, art, and music, we cannot generalize because there were few articles relating to theater, art, and music while there were many articles on product and service reviews. As this is research followed a qualitative research methodology, results and discussions are based on judgement and interpretation of the researcher. While searching for the papers, we came across many studies on online criteria. However, there were few studies on traditional reviews, and scales and measurements used by critics. So, it is clear that many scholars' interest is in researching online review platforms compared to long-established methods. Therefore, we hope that we opened various opportunities for future researchers to position their work and identify potential pathways regarding traditional reviewers and review characteristics. Furthermore, future research may include a representative sample and statistical method to rectify generalizability.

# References

The references in the table are not included due to space limitations. Please contact the authors to receive a copy of the full reference list.